\documentclass[12pt]{article}
\def\Journal#1#2#3#4{{#1}, {\bf #2} #3 (#4)}
\def\be{\begin{equation}}
\def\ee{\end{equation}}
\def\ba{\begin{array}}
\def\ea{\end{array}}
\def\bea{\begin{eqnarray}}
\def\eea{\end{eqnarray}}
\def\bean{\begin{eqnarray*}}
\def\eean{\end{eqnarray*}}
\def\hf{\frac{1}{2}}
\def\i{{\rm i}}
\newcommand{\sca}[2]{\langle #1, #2 \rangle}
\newcommand{\JMP}{{\em J. Math. Phys.} }
\newcommand{\PL}{{\em Phys. Lett.} }

\title{Applications of {\sc Crack} in the Classification of Integrable Systems}
\author{Thomas Wolf\\ Department of Mathematics,
Brock University\\ 500 Glenridge Avenue, St.Catharines, 
Ontario, Canada L2S 3A1\\
email: twolf@brocku.ca}

\begin{document}
\maketitle

\section{Overview}
The talk given by the author at the CRM workshop on Superintegrability
in Sep.\ 2002 and this related paper report on work in two subjects.
One is the collaboration with Vladimir Sokolov and Takayuki Tsuchida
in an effort to classify polynomial integrable vector evolution
equations.  The other is the computer algebra package {\sc Crack}
which did the main computations in solving large bi-linear algebraic
systems. Although originally designed to solve
over-determined systems of partial differential equations
a number of extensions made {\sc Crack} a powerful tool for solving
systems of bi-linear algebraic equations.  Such systems turn up in
many different classification problems some of which were investigated by other
participants of this workshop. In sections \ref{hamilton} and
\ref{non_local} two additional applications are outlined.

In the talk on which this article is based a method to reduce the
length of equations was presented which proved to be useful in solving
the bi-linear algebraic systems. Due to numerous asked questions about
the computer program, a more complete overview is given in the appendix.

\section{The classification of integrable vector evolution equations} \label{one_vector}
The method to use symmetries to classify non-linear evolutionary 1+1
dimensional PDEs is the most productive one and known for some
time (see \cite{SokSh,MikShYam,Fokas,MiShSok}). 

An extension of the simplest approach \cite{IbSh1,Fokas1} 
to the case of so-called vector evolution equations is described in
work with Vladimir Sokolov \cite{SokWol01}.
Examples of vector evolution equations are two different
vector generalisations 
\bea
U_{t} & = & U_{xxx} + \langle U, U\rangle  U_x,  \label{mkdv1} \\
U_{t} & = & U_{xxx} + \langle U,  U\rangle \, U_x + \langle U,  U_x\rangle \, U
\label{mkdv2}
\eea
of the mKdV equation where $U(t,x)$ is an $N$-component vector and 
$\langle \cdot,\cdot\rangle$ stands for the standard scalar product.

In the performed symmetry classification 
we considered equations of the form  
\begin{equation}
U_t=f_n \, U_n+f_{n-1}\, U_{n-1}+\cdots+f_1\,U_1+f_0\,U, 
    \qquad U_i=\frac{\partial^i U}{\partial x^i} .
\label{gensys}
\end{equation}
where $U = (U^1, U^2,\ldots, U^N)$ is an unknown vector of arbitrary 
dimension $N$ and coefficients $f_i$ are polynomials 
of scalar products $\langle U_i,\, U_j\rangle, \, 0\le i\le j \le n.$
 
For more details on possible scalar products, orthogonal symmetry
transformations and generality of $N$ see \cite{SokWol01}.

All such equations were determined that omitted a symmetry of the same form
\[U_{\tau}=g_m \, U_m+g_{m-1}\, U_{m-1}+\cdots+g_1\,U_1+g_0\,U\]
where the differential order $n$ of the equation and the 
order $m$ of the symmetry have selected values.
By taking $f_i, g_j$ to be homogeneous polynomials of the scalar 
products one achieves that the
symmetry condition $U_{t \tau} = U_{\tau t}$ yields an over-determined
system of bi-linear algebraic conditions for the un-determined coefficients 
of both polynomials $f_i$ and $g_i$. 

The number of coefficients of the equation and symmetry is reduced further by
a homogeneity assumption that the differential equations and
symmetries are invariant under the scaling group
$$(x, \ t, \ U)\longrightarrow (a^{-1}x, \ a^{-\mu} t, \ a^{\lambda} U).$$
with one value $\lambda$ for all components of the vector $U$.

Finally, we restrict $\lambda$ based on results for the scalar case
proven in \cite{SW} that a $\lambda$-homogeneous polynomial equation
with $\lambda>0$ may possess a homogeneous polynomial higher symmetry
only if $\lambda \in \{2, 1, {1\over 2}\}.$

The differential orders of the equation and symmetry that have been
investigated are also motivated by the scalar case where infinite
commutative hierarchies have either a lowest order of the equation of
2 and symmetry of order 3 (in the following called type (2,3) like the
Burgers equation) or an equation of order 3 with a symmetry of order 5
(type (3,5) like the Korteweg--de Vries (KdV) equation) or an equation
of order 5 with a symmetry of order 7 (type (5,7) like
the Kaup--Kupershmidt and Sawada--Kotera equations).

For the nine resulting cases shown in table 1
computer programs generated the equations and symmetries, computed the
commutator and formulated and solved the bi-linear system 
automatically (apart from the largest case $\lambda={1 \over 2}$, type
(5,7) where the solution was not fully automatic).

\begin{center}
\scriptsize
\begin{tabular}{|l||c|c|c||c|c|c||c|c|c|} \hline
$\lambda$         &\multicolumn{3}{c||}{2}&\multicolumn{3}{c||}{1}&\multicolumn{3}{c|}{1/2}\\ \hline
orders of (eq,sym)& (2,3) & (3,5) & (5,7) & (2,3) & (3,5) & (5,7) & (2,3) & (3,5) & (5,7) \\ \hline
\# of unknowns (eq,sym,tot)
                  &   -   &   2   &3,7,10 & 2,3,5 &3,9,12 &9,24,33&4,8,12 &8,27,35&27,82,109\\ \hline 
\# of equations   &   -   &   4   &  34   &   5   &  26   &  198  &  21   &  129  &  927  \\ \hline 
total \# of terms &   -   &   4   & 162   &   9   &  121  & 3125  &  80   & 1603  & 52677 \\ \hline 
av.\ \# of terms/equ.\     
                  &   -   &   1   &  4.7  &  1.8  &  4.6  &  15.8  & 3.8  & 12.4  &  56.8 \\ \hline 
time to formulate &   -   &  0.2s &  1.7s &   0s  &  1.3s & 1m 1s  & 1s   &  17s  &13h 12m \\ \hline 
time to solve     &   -   &   0s  &  0.7s &   0s  &  0.5s & 1m 15s & 0.4s &  32s  & 2 days\\ \hline 
solutions         &   -   & - & - & - &(\ref{mkdv1}),(\ref{mkdv2})& - & - &(\ref{ibsh})& - \\ \hline
\end{tabular} \vspace{6pt}\\
\normalsize
Table 1. A classification of single vector equations.
\end{center}

Comments: Times are measured on a 1.7GHz Pentium 4 running a 120MByte
Reduce session under Linux. Nonlinear $\lambda=2$ equations of order 2
do not exist.  All found solutions have already been known: equations
(\ref{mkdv1}), (\ref{mkdv2}) are vector generalizations of the mKdV
equation, and equation 
\be U_{t}=U_{xxx}+3 \langle U, U\rangle U_{xx}+6 \langle U,U_x\rangle U_x
+ 3 \langle U,U\rangle^2 U_x+3\langle U_x, U_x\rangle U
\label{ibsh} \ee
is a vector analogue (reported in \cite{SokWol}) of the
Ibragimov-Shabat-Calogero equation \cite{IbSh1,Calo}.  Solutions of
order (5,7) are only symmetries of found lower order equations
and have therefore not been listed in the table.


\section{NLS Systems with two Vector Unknowns} \label{two_vectors}

More successful has been the investigation of integrable vector NLS-type
systems of the form
\bea
\label{nceqgen}
\cases{ U_{t}=\;\;\,U_{xx}+p_1 U_x+p_2 V_x+p_3 U+p_4 V,\vspace{6pt}
  \cr V_{t}=-V_{xx}+p_5 U_x+p_6 V_x+p_7 U+p_8 V ,\cr} 
\eea 
where $U$ and $V$ are vectors and the coefficients $p_i$ are
$\lambda$-homogeneous polynomials depending on all possible scalar
products of vectors $U,V,U_x,V_x$. For $\lambda=2$ second order systems
can only be linear but for $\lambda=1$ or $1/2$ results are shown in table
2. Just as in the scalar case (see \cite{MikShYam}) a symmetry of the
form \bea
\label{symgen}
\cases{
U_{\tau}=U_{xxx}+q_1 U_{xx}+q_2 V_{xx}+q_3 U_x+q_4 V_x+q_5 U+q_6 V,\vspace{6pt} \cr
V_{\tau}=V_{xxx}+q_7 U_{xx}+q_8 V_{xx}+q_9 U_x+q_{10} V_x+q_{11} U+q_{12} V,
\cr}
\eea
is assumed where the coefficients $q_i$ are $\lambda$-homogeneous 
polynomials of all possible scalar products of $U, V, U_x, V_x, U_{xx}, V_{xx}$.

\begin{center}
\scriptsize
\begin{tabular}{|l||c|c|c||c|c|c||c|c|c|} \hline
$\lambda$         &   1   &  1/2  \\ \hline
orders of (eq,sym)& (2,3) & (2,3) \\ \hline
\# of unknowns (eq,sym,tot)&9,15,24&53,155,208  \\ \hline 
\# of equations   &  78   & 1206  \\ \hline 
total \# of terms & 242   & 28768 \\ \hline 
av.\ \# of terms/equ.\ &  3.1  &  23.8 \\ \hline 
time to formulate & 2.3s  &  1h   \\ \hline 
time to solve     & 2.9s  &26m 15s\\ \hline 
\# of solutions   &   2   &   7   \\ \hline
\end{tabular} \vspace{6pt}\\
\normalsize
Table 2. A classification of NLS-type systems of vector equations.
\end{center}

{\em Results:}

\noindent The 2 solutions for $\lambda=1$ are known (\cite{man,sclan}).
They are special cases (see \cite{SokWol01})
of a generalization of the NLS system by Svinolupov
using a Jordan triple system (in \cite{Sv2}).
For $\lambda={1 \over 2}$ after identifying solutions through
$U \leftrightarrow V, \ t \leftrightarrow -t$
six solutions remain. 
Two of these systems had been known (\cite{japan1}).
For the remaining four systems, Takayuki Tsuchida showed 
C-integrability for two of them and S-integrability for the other two.


\section{Systems with one Scalar and one Vector Unknown} \label{scalar_vectors}
Computations to classify single vector equations that involved an
arbitrary constant vector did not give new results but the possibility
to apply an orthogonal transformation to make the constant vector equal
(1,0,0,...) provided a natural split of the single vector equation
into one for a scalar function $u$ (equal the former component $U^1$) 
and a new vector function $U$ (equal the remaining components $(U^2,U^3,...)$). 
General investigations of systems with one scalar and one vector 
brought surprisingly a rich set of integrable systems. 
Results in the case $\lambda=2$ were analysed by Vladimir Sokolov and
Takayuki Tsuchida. What appears to be a new solution in this class is
the vector generalization 
\bea
\label{cHS}
\cases{
u_t = u_{xxx} + 6 uu_x -12 \sca{U}{U_x}    \vspace{1.5mm}, \cr
U_t = -2 U_{xxx} - 6u U_x.
\cr}
\eea
of the two-component coupled KdV system 
proposed by Hirota and Satsuma \cite{HiSa}. 
A Lax pair representation has been found by 
Takayuki Tsuchida.

Especially the case of $\lambda=1$ with 25
solutions analysed by Takayuki Tsuchida posed a major challenge. 
Not all systems are of interest. Some become triangular by defining $\sca{U}{U}$ 
as a new scalar variable. Others are just the result of splitting
a single vector equation into scalar + vector equations for the
scalar $U^1$ and the vector $(U^2,U^3,\ldots,U^N)$.
From the interesting cases just one should be shown here:
\bea
\label{sol24}
\cases{
u_t = u_{xxx} -6 u^2 u_x + u_x \sca{U}{U} + 2u \sca{U}{U_x} 
      + \sca{U}{U_{xx}} + \sca{U_x}{U_x} 
    \vspace{1.5mm}, \cr
U_t = -2 U_{xxx} - 6 u_{xx} U - 6u_x U_x + 12uu_x U + 6u^2 U_x 
        + \sca{U}{U} U_x 
\cr \qquad \mbox{} -2 \sca{U}{U_x}U.
\cr}
\eea

Its integrability can be established (Tsuchida)
through a change to new variables $w$ and $W$ 
\bea
\label{Miura3}
\cases{
w = u_x + u^2 + \frac{1}{6} \sca{U}{U} \vspace{1.5mm},
\cr
W = U_x + 2uU,
\cr}
\eea
which satisfy the following system:
\bea
\label{cHS2}
\cases{
w_t = w_{xxx} -6ww_x + 2\sca{W}{W_x}
\vspace{1.5mm} ,\cr
W_t = -2 W_{xxx} + 6 w W_x.
\cr}
\eea
This system coincides up to a scaling of variables
with the coupled Hirota--Satsuma 
system (\ref{cHS}) found in the $\lambda=2$ investigation
above.

The Miura-type transformation 
(\ref{Miura3}) is a generalization of 
the one for scalar $U$ in \cite{Wilson,Drinfeld} 
and the one for two-component vector $U$ in \cite{Wu}. 

In the case when $U$ is a scalar variable, one can set
\[u= -\hf (q+r), \quad U = \frac{\sqrt{6}}{2}\i (q-r)\]
and rewrite (\ref{sol24}) as a system of 
symmetrically coupled mKdV equations: 
\bea
\cases{
q_t = \bigl[ -\hf q_{xx} + \frac{3}{2} r_{xx} 
        +3 (q-r) q_x -2 r^3 \bigr]_x
    \vspace{1.5mm}, \cr
r_t = \bigl[ \frac{3}{2} q_{xx}  - \hf r_{xx} - 3(q-r)r_x -2 q^3
        \bigr]_x.
\cr}
\eea
This system is identical to (63) in \cite{Foursov2} 
or (3.22) in \cite{Foursov3}. 
It was found in connection with the Kac--Moody Lie algebras 
and written in the Hamiltonian form about twenty years ago 
(cf.\ the ${\rm C}_2^{(1)}$ case in \cite{Wilson} 
or the ${\rm B}_2^{(1)}$ case in \cite{Drinfeld}). 

The following table gives an overview of systems for one scalar and
one vector found by {\sc Crack} to have higher order symmetries.
Details will be discussed in a future contribution.
\begin{center}
\scriptsize
\begin{tabular}{|l||c|c|c||c|c|c||c|c|c|} \hline
$\lambda$         &\multicolumn{3}{c||}{2}&\multicolumn{3}{c||}{1}&\multicolumn{3}{c|}{1/2}\\ \hline
orders of (eq,sym)         & (2,3) & (2,4) & (3,5) &  (2,3) & (2,4)  & (3,5)  & (2,3)  & (2,4) & (3,5) \\ \hline
\# of unknowns (eq,sym,tot)&5,6,11 &5,12,17&6,17,23&10,21,31&10,39,49&21,74,95&15,36,51&15,79,94&36,164,200 \\ \hline 
\# of equations            &  13   &  26   &  50   &   66   &  123   &  386   &  149   &  313   & 1154  \\ \hline 
total \# of terms          &  34   &  77   & 218   &  341   &  770   & 5000   & 1093   &  3096  & 27695 \\ \hline 
av.\ \# of terms/equ.\     &  2.6  &  2.9  &  4.3  &  5.1   &  6.3   &  13    &  7.3   &   9.9  &  24   \\ \hline 
time to formulate          &  0.5s &   1s  &   5s  &  1.8s  &   5s   &2m 52s  &  8s    &   48s  & 2h 7m \\ \hline 
time to solve              &  0.5s & 0.4s  &  6.5s &   29s  & 1m 58s &5h 47m  &  29s   & 3m 44s & 1 day \\ \hline 
\# of solutions            &   0   &   0   &   4   &   3    &   3    &  25    &   0    &   0    &   2  \\ \hline
\end{tabular} \vspace{12pt}\\

\begin{tabular}{|l||c|c|c||c|c|c||c|c|c|} \hline
$\lambda$ &\multicolumn{3}{c||}{scalar: 1/3, \ vec: 2/3}&\multicolumn{3}{c||}{scalar: 2/3, \ vec: 1/3}\\ \hline
orders of (eq,sym)         & (2,3)  & (2,4)  & (3,5)    &  (2,3) & (2,4)  & (3,5)  \\ \hline
\# of unknowns (eq,sym,tot)&10,24,34&10,54,64&24,115,139&13,22,35&13,66,79&22,126,148 \\ \hline 
\# of equations            &  102   &  215   &  798     &  114   &  276   &  955    \\ \hline 
total \# of terms          &  529   & 1462   & 12694    &  694   & 2435   & 17385   \\ \hline 
av.\ \# of terms/equ.\     &  5.2   &  6.8   &  16      &  6.1   &  8.8   &  18     \\ \hline 
time to formulate          &  3.2s  &  13s   &23m 45s   &  6.3s  &  48s   & 41m 18s   \\ \hline 
time to solve              &   45s  &1m 23s  & 1h 20m   &   22s  & 3m 40s & 1h 7m   \\ \hline 
\# of solutions            &   0    &   0    &   0      &   1    &   2    &  2      \\ \hline
\end{tabular} \vspace{12pt}\\

\normalsize
Table 3. A classification of systems of one scalar + one vector equations.
\end{center}


\section{Classification of Integrable Hamiltonians} \label{hamilton}
In work done together with Olga V.\ Efimovskaya
quadratic Hamiltonians $H$ have been investigated
that have the form
\begin{equation}
H=\langle U, \, A U\rangle+\langle U, \, B V\rangle+
  \langle V, \, C V\rangle+
  \langle N,\,U\rangle+\langle M,\,V\rangle ,
\label{HAM}
\end{equation}
where $U=(U_1 ,U_2, U_3 )$ and $V=(V_1 ,V_2, V_3 )$ are three
dimensional vectors, $A, C$ are symmetrical matrices, $B$ is an arbitrary
matrix and $N, M$ are constant vectors.
Such Hamiltonians are relevant in the dynamics of rigid bodies.

The equations of motion in the rigid body dynamics are defined by a linear 
Poisson bracket of the form
\begin{equation}\label{genlinpuass}
\{Y_{i},\, Y_{j}\}=c^{k}_{ij} \,Y_{k}, \qquad i,j,k=1,\dots,N
\end{equation}
where $c^{k}_{ij}$ are some constants. The evolution of dynamic 
variables  $Y_{1},\dots,Y_{N}$ is defined by the formula
$$
\frac{d}{dt} Y_{i}=\{Y_{i},\, H\},
$$
where $H$ is the Hamiltonian. 

The skew-symmetricity and the Jacobi identity for the linear Poisson bracket 
is equivalent to the fact that $c^{k}_{ij}$ are the structural 
constants of some Lie algebra. It is known that the Hamiltonian 
structure of most cases of rigid body dynamics can be defined by the 
linear Poisson brackets 
\begin{equation} \label{puas}
\{U_{i},U_{j}\}=\varepsilon_{ijk}\,U_{k} , \qquad
\{U_{i},V_{j}\}=\varepsilon_{ijk}\,V_{k} , \qquad
\{V_{i},V_{j}\}=0 , 
\end{equation}
corresponding to the Lie algebra $e(3)$. 
For example,  two classical problems with a Hamiltonian of form
(\ref{HAM}) and the Poisson structure (\ref{puas}) are
\begin{enumerate}
\item the Kirchhoff problem (where $N=0, M=0$ in $H$),
\item the problem of motion of a massive 
      rigid body around a fixed point. In this case $B=C=N=0$.
\end{enumerate}

The bracket  (\ref{puas}) possesses two Casimir functions:
\begin{equation} 
 J_1=V_1^2+V_2^2+V_3^2,
 \qquad J_2=U_1 V_1+U_2 V_2+U_3 V_3.   \label{kazim}
\end{equation}
Therefore to integrate a system on $e(3)$ we need one additional first
integral $I$, functionally independent of $H, J_{1}, J_{2}$. All 
Hamiltonians (\ref{HAM}) admitting an additional polynomial first integral 
of first or second degree are well known.
The main goal of our work is to find all Hamiltonians (\ref{HAM})
that admit an additional polynomial integral of degree 3 or 4. 

There exist two kinds of linear transformations of $U$ and $V$ which 
preserve the Poisson structure (\ref{puas}) and the form of $H$. 
The first kind is defined by 
\begin{equation}
\bar U=T\, U, \qquad   \bar V=T\, V,
\label{tran1}
\end{equation}
where $T$ is an arbitrary constant orthogonal matrix.
The second kind is defined by 
\begin{equation}
\bar U=U+S\, V,
\label{tran2}
\end{equation}
where $S$ is an arbitrary antisymmetric matrix.
With the transformations (\ref{tran1}) we reduce the matrix $A$ to diagonal
form: $A=diag(a_1,a_2,a_3)$.
Transformations (\ref{tran2}) are usually used for a simplification of
matrix $B$.

The generic case $a_1 \neq a_2 \neq a_3 \neq a_1$ has been
investigated in detail and all cases when there exists an additional
polynomial integral are known.
In calculations mentioned below we study the case $a_1=a_2 \ne a_3$.
Using transformation (\ref{tran2}) we can 
have $b_{12}=b_{13}=b_{23}=0$. Subtracting multiples of both casimirs
enables $b_{11}=c_{11}=0$ and gives a Hamiltonian of the form 
\begin{equation}
\begin{array}{lcl}
H & = & U_1^2 + U_2^2 + a_3 U_3^2 + \\
  &   & \\
  &   & 2b_{21}U_2V_1 + 2b_{31}U_3V_1 + 2b_{32}U_3V_2 + 
        2b_{22}U_2V_2 + 2b_{33}U_3V_3 + \\
  &   & \\
  &   & 2c_{12}V_1V_2 + 2c_{13}V_1V_3 + c_{22}V_2^2 + 2c_{23}V_2V_3 +
        c_{33}V_3^2 + \\
  &   & \\
  &   & p_1U_1 + p_2U_2 + p_3U_3 + q_1V_1 + q_2V_2 + q_3V_3.
\end{array}  \label{Hamob}
\end{equation}

We consider Hamiltonians (\ref{Hamob}) that have an additional cubic integral. 
The ansatz for a general first integral $I$ of third degree involves 
80 terms. Together with the 16 unknown constants in (\ref{Hamob}) the
bi-linear algebraic system which results from $\{H,I\}=0$ involves
96 unknowns to be determined. The result of the computation is
summarized in the following theorem.

{\bf Theorem 1}.
\begin{itemize}
\item
 {\it The Hamiltonian} (\ref{Hamob}) 
{\it admits a polynomial integral of third degree iff it has the form}
\begin{equation}
\begin{array}{l}
H= U_{1}^2 + U_{2}^2 + s_{1} U_{3}^2 +
 s_{2} V_{3} U_{3} + s_{3} V_{3}^2 + s_{4} U_{3}+ s_{5} V_{3},
\end{array}  \label{Ham31}
\end{equation}
{\it where}  $s_{i}$ {\it are arbitrary parameters;}
\item
 {\it The Hamiltonian} (\ref{Hamob}) 
{\it admits a polynomial integral of third degree on a special 
level $J_2=0$ of Casimir function (\ref{kazim}) iff either it has the
form (\ref{Ham31}) or the form}
\begin{equation}
\begin{array}{l}
H= U_{1}^2 + U_{2}^2 +4 U_{3}^2 + 4 (s_{1} V_{1} + s_{2} V_{2}) U_{3}
 - (s_{1}^{2} + s_{2}^{2}) V_{3}^2 + \\[4mm]
\qquad s_{3} U_{3} +s_{4} V_{1} + s_{5} V_{2},
\end{array}  \label{Ham32}
\end{equation}
 {\it where}  $s_{i}$ {\it are arbitrary parameters.}
\end{itemize}
 
Hamiltonian (\ref{Hamob}) is a trivial generalization of the Hamiltonian for 
the Lagrange and Kirchhoff classical integrable cases. Actually 
Hamiltonian (\ref{Hamob}) admits not only a third degree but  
also an additional first degree integral $I=U_3$.     
If $s_1=s_2=0$  then Hamiltonian (\ref{Ham32}) describes the so-called 
Goryachev-Chaplygin case in the problem of motion of a rigid body around 
a fixed point. The integrability of the general Hamiltonian (\ref{Hamob}) 
has been recently established in {\cite{soktsig}}.


\section{Non-local 2+1 Dimensional Equations}  \label{non_local}
The computation decribed in this section solves only a first special
case of a wider problem. We still show it as it gives an example of
how even non-local 2+1 dimensional classification problems can be
reduced to the solution of bi-linear algebraic systems for which 
{\sc Crack} can be useful.

The Kadomtsev--Petviashvili equation
\[ u_{tx} = -(6uu_x+u_{xxx})_x+u_{yy} \]
can also be written as
\[ u_t = -(6uu_x+u_{xxx})+\Delta^2u_x \]
with $\Delta = D_x^{-1}D_y$. 
A.\ Mikhailov and R.I.\ Yamilov observed in \cite{MY98} 
that all known integrable
2+1 - dim equations can be written as
\[ u_t = \mbox{expression in} \; \Delta, D_x \; \mbox{and} \;  u.\]

Based on this idea A.\ Mikhailov, V.V.\ Sokolov and
R.\ Hernandez Heredero did work at classifying KdV-type
integro-differential equations of the form
\begin{eqnarray*}
u_t&=&a u_{xxx} + b \Delta(u_{xxx}) + c \Delta^2(u_{xxx}) + ... \\
   & &+ e \Delta^{-3}(u_{xxx}) + \mbox{terms of lower $x$-order} 
\end{eqnarray*}
which have symmetries of the same form.

The special ansatz for equation plus symmetry that has been investigated
with {\sc Crack} is
\begin{eqnarray}
u_t&=&u_{xxx} + b \Delta(u_{xxx}) + c \Delta^2(u_{xxx})  \nonumber \\
   & &+ 0\cdot\Delta^3(u_{xxx}) + e \Delta^{-1}(u_{xxx}) + 
       \mbox{lower order} \label{nonloc1} \\
u_\tau&=&0\cdot u_{xxx} + b'\Delta (u_{xxx}) + c'\Delta^2(u_{xxx})
   \nonumber \\
   & &+ \Delta^3(u_{xxx}) + e'\Delta^{-1}(u_{xxx}) + \mbox{lower
   order} \nonumber
\end{eqnarray}
with two coefficients equal one and two zero coefficients due to
suitable linear combinations of both equations.  The lower order terms
can be of $x$-order 2 at most, linear or quadratic in $u$, and with
$\Delta$ restricted as above.

For this ansatz, which is the simplest in this class,
the condition $u_{t\tau}-u_{\tau t} \equiv 0$ provides already 2865 bilinear
algebraic conditions for the 70 unknowns $b,c,\ldots$ and 70 unknowns
$b',c',\ldots $ . 

\noindent According to CRACK only 2 solutions exist.
One is the Boiti, Leon, Manna, Pempinelli equation
$$u_t=u_{xxx}+\alpha D_x(u\Delta^{-1}(u))$$
with symmetry
$$u_\tau=\Delta^3(u_{xxx})+D_x(\Delta(u\Delta(u)))$$
The second solution is the same for $u \rightarrow \Delta u$.

Other known 2-dimensional integrable equations, like 
KdV in the form $u_t=\Delta u_{xxx}+4 u\Delta u_x+2 u_x \Delta u $
which is a symmetry of the usual KdV equation 
would not have terms $u_{xxx}$ and $\Delta^3(u_{xxx})$ as it was
required in the special ansatz (\ref{nonloc1}).


\section*{Appendix: A short description of {\sc Crack}}
Any identifiers or numbers in curled brackets \{ \} provided at the
end or within the following paragraphs refer to key-words, file
names, module numbers or flags which can be looked up in the {\sc Crack} manual
{\tt crack.tex} (see below under `Availability') or even be searched in the
source code if needed.

\begin{description}
\item[Philosophy:] The program {\sc Crack} is a computer algebra
  package written in REDUCE for the solution of over-determined
  systems of algebraic, ordinary or partial differential equations
  with at most polynomial non-linearity. It was originally developed to
  run automatically and effort was taken for the program to decide which
  computational steps are to be done next with a choice between
  integrations, separations, substitutions and investigation of
  integrability conditions. It is known from hand computations that
  the right sequence of operations with exactly the right equations
  at the right time is often crucial to avoid an explosion of the length
  of expressions. This statement keeps its truth for the computerized
  solution of systems of equations as they become more complex. As a
  consequence more and more interactive access has been provided to
  inspect data, to specify how to proceed with the computation and how to
  control it. This allows the human intervention in critical stages of
  the computations. {\em \{off batch\_mode\} }
\item[General Structure:] A problem consists of a system of equations
  and a set of inequalities. With each equation are associated a short
  name and numerous data, like size, which functions, derivatives and
  variables occur but also which investigations have already been done
  with this equation and which not in order to avoid unnecessary
  duplication of work.  These data are constantly updated if the
  equation is modified in any way.
  
  A set of about 30 modules is available to integrate, substitute,
  decouple, ... equations. A complete list can be inspected in
  interactive mode with the command {\em p2}, each operation is listed
  with the number it is called. All modules can be called
  interactively or automatically. Automatic computation is organized
  by a priority list of modules (each represented by a number) where
  modules are invoked in the order they appear in the priority list,
  each module trying to find equations in the system it can be applied
  to.  If a module is not successful then the next module in the list
  is tried, if any one is successful then execution starts again at
  the beginning of the priority list. {\em \{ prog\_list\_,
    default\_proc\_list\_, full\_proc\_list\_\}}

  Because each module has access to all the data, it is enough to call
  a module by its number. For example, the input of the number 2 in
  interactive mode will start the direct separation module (see below)
  to look for a directly separable equation and will split it.
  
\item[Modules:] The following modules are represented by numbers in
  the priority list. Each module can appear with modifications under
  different numbers.  For example, integration is available under 7,
  24 and 25.  Here 7 encodes an integrations of short equations
  $0=\partial^n f/\partial x^n$. 7 has highest priority of the three
  integrations. 24 encodes the integration of an equation that leads
  to the substitution of a function and 25 refers to any integration
  and has lowest priority.
  \begin{description}
  \item[Integration and Separation:]
    An early feature in the development of the package {\sc Crack}
    was the ability to integrate exact differential equations and
    some generalizations of them (see \cite{Wol99e}). As a consequence
    of integrations an increasing number of functions of fewer
    variables is introduced which sooner or later produces equations with some
    independent variables occuring only explicitly and not as
    variables in functions. Such equations are splitted by the
    integration module.
  \item[Substitutions:]
    Substitutions can have a dramatic effect on the size and
    complexity of systems. Therefore it is possible to have them not
    only done automatically but also controlled tightly,
    either by specifying exactly which unknown should be substituted
    where using which equation, or by picking a substitution out of a list
    of substitutions offered by the program {\em \{cs\} }. Substitutions to be
    performed automatically can be controlled with a number of filters,
    for example, by
    \begin{itemize}
    \item limiting the size of the equation to be used for
    substitution, {\em \{length\_limit\} }
    \item limiting the size of equations in which the
    substitution is to be done, {\em \{pdelimit\} }
    \item allowing only linear equations to 
    be used for substitutions, {\em \{lin\_subst\} }
    \item allowing equations to increase in size only up 
    to some factor in order for a substitution to be performed 
    in that equation, {\em \{cost\_limit\} }
    \item allowing a substitution for a function through an expression
    only if that expression involves exclusively functions of fewer
    variables, {\em \{less\_vars\} }
    \item allowing substitutions only that do not lead to a case distinction
    coefficient = 0 or not,
    \item specifying whether extra effort should be spent to identify the
    substitution with the lowest bound on growth of the full system.
    {\em \{min\_growth\} } 
    \end{itemize}
    Substitution types are represented by different numbers depending
    on the subset of the above filters to be used.  If a substitution
    type is to be done automatically then from all possible
    substitutions passing all filters of this type that substitution
    is selected that leads to no sub-cases (if available) and that
    uses the shortest equation.
  \item[Factorization:]
    It is very common that big algebraic systems contain equations that
    can be factorized. Factorizing an equation and setting the factors
    individually to zero simplifies the whole task because factors are
    simpler expressions than the whole equation and set to zero they may
    lead to substitutions and thereby further simplifications. The
    downside is that if problems with, say 100 unknowns, need 40
    case-distinctions in order to be able to solve automatically for the
    remaining 60 unknowns then this would require $2^{40}$ cases to be
    investigated which is impractical. The problem is to find the right
    balance, between delaying case-distinctions in order not to generate
    too many cases and on the other hand introducing case distinctions
    as early as necessary in order to simplify the system. This
    simplification may be necessary to solve the system but in any
    case it will speed up its solution (although at the price of having
    to solve a simplified system at least twice, depending on the number
    of factors).
    
    For large systems with many factorizable equations the careful
    selection of the next equation to be factorized is important to
    gain the most from each factorization and to succeed with as few
    as possible factorizations.  Criteria which give factors and
    therefore equations a higher priority are
    \begin{itemize}
    \item the number of equations in which this factor occurs,
    \item if the factor is a single unknown function or constant, 
      then the number of times this
      unknown turns up in the whole system,
    \item the total degree of the factor,
    \item the number of factors of an equation,
    \item and others.
    \end{itemize}
    It also matters in which order the factors are set to zero. For
    example, the equation  $0=ab$ can be used to split into the 2
    cases: $1.\ a=0, \ 2.\ a\neq 0, b=0$ or to split into the 2 cases
    $1.\ b=0, \ 2.\ b\neq 0, a=0$. If one of the 2 factors, say $b$, involves
    functions which occur only linearly then this property is to be
    preserved and these functions should be substituted as such
    substitutions preserve their linearity. But to have many such
    substitutions available, it is useful to know of many non-linearly
    occuring functions to be non-zero as they occur as coefficients of
    the linearly occuring functions. In the above situation it is
    therefore better to do the first splitting $1.\ a=0, \ 2.\ a\neq 0, b=0$
    because $a\neq 0$ will be more useful for substitutions of linear
    functions than $b\neq 0$ would be. 

    An exception of this plausible
    rule occurs towards the end of all the substitutions of all the
    linearly occuring $b_i$ when some $b_i$ are an overall factor to
    many equations. If one would then set, say, $b_{22}=0$ as the second
    case in a factorization, the first case would generate as subcases
    factorizations of other equations where $b_{22}=0$ would be the second case
    again and so on. To avoid this one should investigate $b_{22}=0$ as the
    first case in the first factorization.

    The only purpose of that little thought experiment was to show
    that simple questions, like which factored equation should 
    be used first for case-distinctions and in which order to set factors to
    zero can already be difficult to answer in general.
  \item[Elimination (Gr\"{o}bner Basis) Steps:] 
    To increase safety and avoid excessive expression swell 
    one can apart from the normal call {\em \{30\} } 
    request to do Gr\"{o}bner basis computation
    steps only if they are simplification steps replacing an equation by
    a shorter equation. {\em \{27\} } 

    In a different version only steps are performed in which equations
    are included which do not contain more than 3 unknowns. This helps
    to focus on steps which are more likely to solve small sub-systems
    with readily available simple results. {\em \{57\} } 

    Often the computationally cheapest way to obtain a consistent
    (involutive) system of equations implies to change the ordering
    during the computation. This is the case when substitutions 
    of functions are performed which are not ranked highest in
    a lexicographical ordering of functions. But {\sc Crack} also
    offers an interactive way to
    \begin{itemize}
    \item change the lexicographical ordering of variables, {\em \{ov\} } 
    \item change the lexicographical ordering of functions, {\em \{of\} } 
    \item give the differential order of derivatives a higher or lower
      priority in the total ordering than the lexicographical ordering
      of functions, {\em \{og\} } 
    \item give either the total differential order of a derivative of a
      function a higher priority than the lexicographical ordering of
      the derivative of that function or to take the lexicographical
      ordering of derivatives as the only criterium. {\em \{of\} } 
    \end{itemize}
  \item[Solution of an under-determined differential equation:]
    When solving an over-determined system of linear differential
    equations where the general solution involves free functions, then
    in the last computational step often a single equation for more
    than one function remains to be solved. Examples are the computation
    of symmetries and conservation laws of non-linear differential
    equations which are linearizable. In {\sc Crack} two procedures are
    available, one for under-determined linear ODEs {\em \{22\} } 
    and one for linear PDEs, {\em \{23\} } 
    both with non-constant coefficients.
  \item[Indirect Separation:]
    Due to integrations new functions of fewer variables are
    introduced. Substituting functions may lead to equations where no
    function depends on all variables but all variables appear as
    variables to unknown functions, e.g.\ $0=f(x)+g(y)$ although
    usually much more complicated with 10 or 20 independent variables
    and many functions depending on different combinations of these
    variables. Because no variable occurs only explicitly, direct
    separations mentioned above are not possible.
    Two different algorithms, one for linear indirectly separable
    equations {\em \{10, 26\} } 
    and one for non-linear directly separable equations {\em \{48\} } 
    provide systematic ways of dealing with such equations.
    
    Indirectly separable equations always result when an equation is
    integrated with respect to different variables, like $0=f_{xy}$ to
    $f=g(x)+h(y)$ and a function, here $f(x,y)$, is substituted.
  \item[Function and variable transformations:]
    In the interactive mode one can specify a transformation of the whole
    problem in which old functions and variables are expressed in a mix
    of new functions and variables.
  \end{description}
  We conclude the listing of modules and continue with other aspects
  of {\sc Crack}..
\item[Exploiting Bi-linearity:]
  In bi-linear algebraic problems we have 2 sets of variables
  $a_1,..,a_m$ and $b_1,..,b_n$ such that all equations have
  the form $0=\sum_{k=1}^l \gamma_k a_{i_k}b_{j_k}, \ \gamma_k \in G$.  
  Although the
  problem is linear in the $a_i$ and linear in the $b_j$ it still is a
  non-linear problem. A guideline which helps keeping the structure of
  the system during computation relatively simple is to preserve the
  linearity of either the $a_i$ or the $b_j$ as long as possible. In
  classification problems of integrable systems
  the ansatz for the symmetry/first integral
  involves usually more terms and therefore more constants (called
  $b_j$ in applications of {\sc Crack}) than the ansatz for the
  integrable system (with constants $a_i$).  A good strategy therefore
  is to keep the system linear in the $b_j$ during the computation,
  i.e.\ to
  \begin{itemize}
  \item substitute only a $b_j$ in terms of $a_i, b_k$, or an $a_i$
    in terms of an $a_k$ but not an $a_i$ in terms of any $b_k$,
  \item do elimination steps for any $b_j$ or for an
    $a_i$ if the involved equations do not contain any $b_k$,
  \end{itemize}
  The proposed measures are effective not only for algebraic problems
  but for ODEs/PDEs too (i.e.\ to preserve linearity of a sub-set of
  functions as long as possible).
  {\em \{flin\_\} } 
\item[Flexible Process Control:]
  Different types of over-determined systems are more or less suited
  for an automatic solution. With the currrent version (2002)
  it is relatively save to try solving large bi-linear algebraic problems
  automatically. Another well suited area concerns over-determined systems
  of linear PDEs. In contrast, non-linear systems of PDEs most likely
  require a more tight interactive control. Different modes of
  operation are possible. One can 
  \begin{itemize}
  \item perform one {\em\{a\}} or more computational steps {\em\{g\}} 
    automatically, each step trying modules in the order defined by the
    current priority list {\em \{p1\} } until one module succeeds in
    its purpose,
  \item 
    perform one module a specific number of times or as long as it is
    successful, {\em \{l\} }
  \item
    set a time limit until which the program should run automatically,
    {\em \{time\_limit, limit\_time\} } 
  \item 
    interrupt an on-going automatic computation and continue the
    computation interactively, {\em\{\_stop\_\} }
  \item  
    arrange that the priority list of modules changes
    at a certain point in the computation when the system of equations
    has changed its character,
  \item induce a case distinction whether a user-given expression 
    is zero or not. {\em \{44\} } 
  \end{itemize}
  Apart from flexible control over what kind of steps to be done, the
  steps themselves can be controlled more or less too, e.g.\ whether
  equations are selected by the module or the user.

  Highest priority in the priority list have so-called {\em to-do}
  steps. The list of to-do steps is usually empty but can be filled by
  any successful step if it requires another specific step to follow
  instantly. For example, if a very simple equation $0=f_x$ is
  integrated then the substitution of $f$ should follow straight away, even
  if substitutions would have a low priority according to the 
  current priority list. 
\item[Total Data Control:]
  To make wise decisions of how to continue
  the computation in an interactive session one needs tools to inspect large systems of
  equations. 
  Helpful commands in {\sc Crack} print
  \begin{itemize}
  \item equations, inequalities, functions and variables, {\em \{e, pi, f\} } 
  \item the occurence of all derivatives of selected functions in any
    equation, {\em \{v\} } 
  \item a statistics summary of the equations of the system, {\em \{s\} } 
  \item a matrix display of occurences of unknowns in all equations, {\em \{pd\} } 
  \item the value of any LISP variable, {\em \{pv\} } 
  \item the value of algebraic expressions that can be specified using
        equation names \\
        (e.g.\ {\tt coeffn(e\_5,df(f,x,y),2)}), {\em \{pe\} } 
  \item not under-determined subsystems. {\em \{ss\} } 
  \end{itemize}
\item[Safety:]
  When working on large problems it may come to a stage where computational
  steps are necessary, like substitution, which are risky in the sense
  that they may simplify the problem but also complicate it by
  increasing its size. To avoid this risk a few safety features have
  been implemented. 
  \begin{itemize}
  \item At any time during the computation one can save a backup of
    the complete current situation in a file and also load a backup. {\em \{sb, rb\} } 
  \item All key strokes are automatically recorded in a list and are
    available after each interactive step, or when the computation has
    finished.  This list can be fed into {\sc Crack} at the beginning
    of a new computation so that the same operations are performed
    automatically that were performed interactively before. The
    purpose is to be able to do an interactive exploration first and
    to repeat it afterwards automatically without having to note with
    pen or pencil all steps that had been done.  {\em \{history\_, old\_history\} } 
  \item During an automatic computation the program might start a
    computational step which turns out to take far too long. It would
    be better to stop this computation and try something else instead.
    But in computer algebra with lots of global variables involved it
    is not straight forward to stop a computation in the middle of it.
    If one would use time as a criterion then it could happen that
    time is up during a garbage collection which to stop would
    be deadly for the session. {\sc Crack} allows to
    set a limit of garbage collections for any one of those
    computations that have the potential to last forever, like
    algebraic factorizations of large expressions.
    With such an arrangement an automatic computation
    can not get stuck anymore due to lengthy factorizations, searches
    for length reductions or elimination steps.
    {\em \{max\_gc\_elimin, max\_gc\_fac, max\_gc\_red\_len, 
    max\_gc\_short, max\_gc\_ss\} } 
  \item Due to a recent (April 2002) initiative of Winfried Neun the
    parallel version of the computer algebra system REDUCE has been
    re-activated and is running on the Beowulf cluster at Brock
    University \cite{MelNeun}. This allows conveniently (with a
    2-letter command) to duplicate the current status of a 
    {\sc Crack} computation to another computer, to try out there different
    operations (e.g.\ risky ones) until a viable way to continue the
    computation is found without endangering the original
    session. {\em \{pp\} } 
  \end{itemize}
\item[Managing Solutions:]
  Non-linear problems can have many solutions. The number of solutions
  found by {\sc Crack} can even be higher because to make progress
  {\sc Crack} may have factorized an equation and considered the two
  cases $a=0$ and $a \neq 0$ whereas solutions in both cases could be
  merged to only one solution without any restriction for $a$. This merging
  of solutions can be accomplished with a separate program {\tt
  merge\_sol()} after the computation.

  Another form of post-processing is the production of 
  a web page for each solution, like \\
  \verb+http://lie.math.brocku.ca/twolf/bl/v1l05o35-s1.html+ .

  If in the solution of over-determined differential equations the
  program performs integrations of equations before the differential
  Gr\"{o}bner basis was computed then in the final solution there may
  be redundant constants or functions of integration. 
  Redundant constants or functions in a solution are not an
  error but they makes solutions appear unnecessarily complicated.
  In a postprocessing step these functions and constants can be eliminated.
  {\em \{adjust\_fnc, drop\_const(), dropredundant()\} } 

\item[Parallelization:]
  The availability of a parallel version of {\sc Crack} was mentioned
  above allowing to try out different ways to continue an ongoing
  computation. A different possibility to make use of a cluster of
  computers with {\sc Crack} is, to export automatically the
  investigation of sub-cases and sub-sub-cases to different computers
  to be solved in parallel.
  
  It was explained above how factorizations may be necessary to make
  any progress but also their potential of exploding the time requirements.
  By running the computation on a cluster and
  being able to solve many more cases one can give factorizations a
  higher priority and capitalize on the benefit of factorizations,
  i.e.\ the simplification of the problem.

\item[Relationship to Gr\"{o}bner Basis Algorithms:]
  For systems of equations in which the unknown constants or functions
  turn up only polynomially a well known method is able to check the
  consistency of the system. For algebraic systems this is the
  Gr\"{o}bner Basis Method and for systems of differential equations
  this is the differential Gr\"{o}bner Basis method. To guarantee the
  method to terminate a total ordering of unknowns and their
  derivatives has to be introduced. This ordering determines which
  highest powers of unknowns are to be eliminated next or which
  highest order derivatives have to be eliminated next using integrability
  conditions. Often such eliminations lead to exponential growth of
  the generated equations. In the package {\sc Crack} such
  computations are executed with only a low priority. A higher
  priority have operations which reduce the length of equations,
  irrespective of any orderings. Violating any ordering a finite
  number of times still guarantees a finite algorithm. The potential
  gain is large as described next.

\item[Length Reduction of Equations:]
  An algorithm designed originally to length-reduce differential
  equations proved to be essential in length reducing systems of
  bi-linear algebraic equations or homogeneous equations which resulted
  from bi-linear equations during the solution process. 

  The aim of the method is to find out whether one equation $0=E_1$ can
  length reduce another one $0=E_2$ by replacing $E_2$ through an
  appropriate linear combination $\alpha E_1 - \beta E_2,\ \ \beta \neq
  0$. To find $\alpha, \beta$ one can divide each term of $E_2$ through
  each term of $E_1$ and count how often each quotient occurs.  If a
  quotient $\alpha/\beta$ occurs $m$ times then $\alpha E_1 - \beta E_2$
  will have $\leq n_1+n_2-2m$ terms because $2m$ terms will cancel each
  other. A length reduction is found if 
  $n_1+n_2-2m\leq \ \mbox{max}(n_1,n_2)$. 
  The method becomes efficient after a few algorithmic refinements discussed
  in \cite{Wol99b}. Length reduced equations
  \begin{itemize}
  \item are more likely to length reduce other equations,
  \item are much more likely to be factorizable,
  \item are more suited for substitutions as the substitution
    induces less growth of the whole systems and introduces fewer new
    occurences of functions in equations,
  \item are more likely to be integrable by being exact or being an ODE
    if the system consists of differential equations,
  \item involve on average fewer unknowns and make the whole system more
    sparse. This sparseness can be used to plan better a sequence of
    eliminations.
  \end{itemize}
\item[Customization:]
  The addition of new modules to perform new specialized computations is easy.
  The new module only has to accept as input the system of equations,
  list of inequalities, list of unknowns to be computed and provide
  output in a similar form. The module name has to be added to a
  list of all modules and a one line description has to be added to a
  list of descriptions.
  This makes it easy for users to add special techniques for the solution of
  systems with extra structure. A dummy template module {\em \{58\} } 
  is already added and has only to be filled with content.
\item[Debugging:]
  A feature useful mainly for debugging is that in the middle of an
  ongoing interactive computation the program can be changed by
  loading a different version of {\sc Crack} procedures. Thus one
  could advance quickly close to the point in the execution where an
  error occurs, load a version of the faulty procedure that gives
  extensive output and watch how the fault happens before fixing it.
  
  The possibility to interrupt REDUCE itself temporarily and to
  inspect the underlying LISP environment {\em \{br\} } 
  or to execute LISP commands and to continue with the {\sc Crack}
  session afterwards {\em \{pc\} } 
  led to a few improvements and fixes in REDUCE itself.
\item[Availability:]
  The package {\sc Crack} including manual can be downloaded for free from \newline
  \verb+http://lie.math.brocku.ca/twolf/crack/+. It is requested
  to cite this paper in a publication if {\sc Crack} has been used
  for any computations that contributed to that publication.
\end{description}


\end{document}